\documentclass[aps,prd,twocolumn,superscriptaddress,showpacs]{revtex4}
\usepackage[dvips]{graphicx}
\usepackage{amsmath,amssymb,times}

\newcommand{\bequ}{\begin{equation}}
\newcommand{\eequ}{\end{equation}}
\newcommand{\bea}{\begin{eqnarray}}
\newcommand{\eea}{\end{eqnarray}}

\renewcommand{\a}{\alpha}

\DeclareSymbolFont{boldletters}{OML}{cmm} {b}{it}
\DeclareSymbolFontAlphabet{\mathbit}{boldletters}
\DeclareMathSymbol{\alpha}{\mathalpha}{letters}{"0B}
\DeclareMathSymbol{\beta}{\mathalpha}{letters}{"0C}
\DeclareMathSymbol{\gamma}{\mathalpha}{letters}{"0D}
\DeclareMathSymbol{\delta}{\mathalpha}{letters}{"0E}
\DeclareMathSymbol{\epsilon}{\mathalpha}{letters}{"0F}
\DeclareMathSymbol{\zeta}{\mathalpha}{letters}{"10}
\DeclareMathSymbol{\eta}{\mathalpha}{letters}{"11}
\DeclareMathSymbol{\theta}{\mathalpha}{letters}{"12}
\DeclareMathSymbol{\iota}{\mathalpha}{letters}{"13}
\DeclareMathSymbol{\kappa}{\mathalpha}{letters}{"14}
\DeclareMathSymbol{\lambda}{\mathalpha}{letters}{"15}
\DeclareMathSymbol{\mu}{\mathalpha}{letters}{"16}
\DeclareMathSymbol{\nu}{\mathalpha}{letters}{"17}
\DeclareMathSymbol{\xi}{\mathalpha}{letters}{"18}
\DeclareMathSymbol{\pi}{\mathalpha}{letters}{"19}
\DeclareMathSymbol{\rho}{\mathalpha}{letters}{"1A}
\DeclareMathSymbol{\sigma}{\mathalpha}{letters}{"1B}
\DeclareMathSymbol{\tau}{\mathalpha}{letters}{"1C}
\DeclareMathSymbol{\upsilon}{\mathalpha}{letters}{"1D}
\DeclareMathSymbol{\phi}{\mathalpha}{letters}{"1E}
\DeclareMathSymbol{\chi}{\mathalpha}{letters}{"1F}
\DeclareMathSymbol{\psi}{\mathalpha}{letters}{"20}
\DeclareMathSymbol{\omega}{\mathalpha}{letters}{"21}
\DeclareMathSymbol{\varepsilon}{\mathalpha}{letters}{"22}
\DeclareMathSymbol{\vartheta}{\mathalpha}{letters}{"23}
\DeclareMathSymbol{\varpi}{\mathalpha}{letters}{"24}
\DeclareMathSymbol{\varrho}{\mathalpha}{letters}{"25}
\DeclareMathSymbol{\varsigma}{\mathalpha}{letters}{"26}
\DeclareMathSymbol{\varphi}{\mathalpha}{letters}{"27}
\DeclareMathSymbol{\Gamma}{\mathalpha}{letters}{"00}
\DeclareMathSymbol{\Delta}{\mathalpha}{letters}{"01}
\DeclareMathSymbol{\Theta}{\mathalpha}{letters}{"02}
\DeclareMathSymbol{\Lambda}{\mathalpha}{letters}{"03}
\DeclareMathSymbol{\Xi}{\mathalpha}{letters}{"04}
\DeclareMathSymbol{\Pi}{\mathalpha}{letters}{"05}
\DeclareMathSymbol{\Sigma}{\mathalpha}{letters}{"06}
\DeclareMathSymbol{\Upsilon}{\mathalpha}{letters}{"07}
\DeclareMathSymbol{\Phi}{\mathalpha}{letters}{"08}
\DeclareMathSymbol{\Psi}{\mathalpha}{letters}{"09}
\DeclareMathSymbol{\Omega}{\mathalpha}{letters}{"0A}

\begin{document}
\preprint{SAGA-HE-267}
\title{Violations of parity and charge conjugation in the $\theta$ vacuum 
with imaginary chemical potential}
\author{Hiroaki Kouno}
\email[]{kounoh@cc.saga-u.ac.jp}
\affiliation{Department of Physics, Saga University,
             Saga 840-8502, Japan}

\author{Yuji Sakai}
\email[]{sakai@phys.kyushu-u.ac.jp}
\affiliation{Department of Physics, Graduate School of Sciences, Kyushu University,
             Fukuoka 812-8581, Japan}

\author{Takahiro Sasaki}
\email[]{sasaki@phys.kyushu-u.ac.jp}
\affiliation{Department of Physics, Graduate School of Sciences, Kyushu University,
             Fukuoka 812-8581, Japan}

\author{Kouji Kashiwa}
\email[]{kashiwa@phys.kyushu-u.ac.jp}
\affiliation{Department of Physics, Graduate School of Sciences, Kyushu University,
             Fukuoka 812-8581, Japan}

\author{Masanobu Yahiro}
\email[]{yahiro@phys.kyushu-u.ac.jp}
\affiliation{Department of Physics, Graduate School of Sciences, Kyushu University,
             Fukuoka 812-8581, Japan}

\date{\today}

\begin{abstract}
Charge conjugation (C) and Parity (P) are exact symmetries 
at $\theta =\pi$ and $\Theta \equiv \mu/(iT)=\pi$, 
where $\theta$ is the parameter of the so-called $\theta$-vacuum, 
$\mu$ is the imaginary quark-number chemical potential and $T$ is 
the temperature. Spontaneous breakings of these discrete symmetries 
are investigated by 
the the Polyakov-loop extended Nambu--Jona-Lasinio (PNJL) model.  
At zero $T$, 
P symmetry is spontaneously broken while C symmetry is conserved. 
As $T$ increases, P symmetry is restored just after C symmetry is 
spontaneously broken, so that either P or C symmetry or 
both the symmetries are spontaneously broken for any $T$.
The chiral-symmetry restoration and the deconfinement transition 
at $\theta =\Theta =0$ 
are remnants of the P restoration and the C breaking 
at $\theta =\Theta =\pi$, respectively. 
\end{abstract}

\maketitle

\section{Introduction}
\label{Introduction}

Violations of parity (P), charge conjugation (C) and 
charge-parity symmetries (CP) in strong interaction are 
one of important subjects in particle and nuclear physics. 
For example, the strong CP problem is a long-standing puzzle; 
see for example Ref.~\cite{Vicari} for a review of this problem. 
Lorentz and gauge invariance allow the Quantum Chromodynamics (QCD) 
action to have a term 
\bea
   {\cal L}_{\theta}=\theta \frac{g^2}{64\pi^2}\epsilon^{\mu\nu\sigma\rho}
   F^{a}_{\mu\nu}F^{a}_{\sigma\rho} 
\eea
of the topological charge, where 
$F^{a}_{\mu\nu}$ is the field strength of gluon. The parameter  
$\theta$ can take any arbitrary value between $-\pi$ and $\pi$, where 
$\theta=-\pi$ is identical with $\theta=\pi$. 
Nevertheless, experiment indicates 
$|\theta| < 3 \times 10^{-10}$~\cite{Baker}. 
Since $\theta$ is P-odd (CP-odd), 
P (CP) is then preserved for $\theta=0$ and $\pm \pi$, but 
explicitly broken for other $\theta$. 
Why is $\theta$ so small ? This is the so-called strong CP problem. 

For zero temperature ($T$) and zero quark-chemical potential ($\mu$), 
P is conserved at $\theta=0$, as Vafa and Witten showed~\cite{VW}. 
Meanwhile, P is spontaneously broken 
at $\theta =\pi$, as Dashen~\cite{Dashen} 
and Witten~\cite{Witten} pointed out. 
This is the so-called Dashen phenomena. 
Since the spontaneous P violation is a nonperturbative phenomenon, 
the phenomenon was so far studied mainly with the effective model 
such as the chiral perturbation theory~\cite{VV,Smilga,Tytgat,ALS,Creutz,MZ}.

For $T$ higher than the QCD scale $\Lambda_{\mathrm QCD}$, 
there is a possibility that a finite $\theta$, depending 
on spacetime coordinates $(t,x)$, is effectively 
induced~\cite{Fukushima:2008xe}, 
since sphalerons are so activated as to jump over the potential 
barrier between the different degenerate ground states~\cite{MMS}. 
If so, P symmetry can be violated locally in high-energy heavy ion collisions. 
This effective $\theta (t,x)$ deviates 
the total number of particles plus antiparticles with right-handed 
$helicity$ from that with left-handed $helicity$. 
The magnetic field, formed in the early stage of heavy-ion collision, 
will lift the degeneracy in spin depending on the charge of particle. 
As a consequence of this fact, 
an electromagnetic current is generated along the magnetic field, 
since particles with right-handed helicity moves opposite 
to antiparticles with right-handed helicity. 
This is the so-called chiral magnetic effect 
(CME)~\cite{Kharzeev,FKW,Fukushima3}. 
CME may explain the charge separations 
observed in the recent STAR results~\cite{Abelev}. 
Thus, theoretical study on the thermal system with nonzero $\theta$ is 
interesting.

For finite $\mu$, the QCD action has a term 
\bea
   {\cal L}_{\mu}=\frac{\mu}{T} {\bar q}\gamma_0 q 
\eea
of the baryon-number charge, where $q$ is the quark field. 
When $\mu$ is pure imaginary, i.e. $\mu=i\Theta T$, ${\cal L}_{\mu}$ 
has a mathematical structure similar to ${\cal L}_{\theta}$, 
if the baryon number is conserved.  
The dimensionless chemical potential $\Theta$ can vary from $-\pi$ to $\pi$, 
where $\Theta=-\pi$ is identical with $\Theta=\pi$. 
Since $\Theta$ is a C-odd quantity, 
C is an exact symmetry at $\Theta=0$ and $\pm \pi$, 
but not at other $\Theta$. 
Thus, C violation induced by finite $\Theta$ is analogous in principle 
to P violation induced by finite $\theta$. 

At imaginary $\mu$, QCD has a periodicity of $2\pi/3$ in $\Theta$. 
This periodicity was found by Roberge and Weiss with 
perturbative QCD  for high $T$ and 
strong-coupling QCD~\cite{RW} for low $T$. This Roberge-Weiss (RW) 
periodicity was confirmed by lattice QCD (LQCD)~\cite{FP,Elia,Chen34,Chen,D'Elia-iso,Cea,D'Elia-3,FP2010,Nagata,Takaishi}. 
At higher temperature, there exist three 
$\mathbb{Z}_3$ vacua. 
As $\Theta$ increases from $-\pi$ to $\pi$, 
the three vacua emerge one by one. As a consequence of this mechanism, 
three first-order phase-transitions appear at $\Theta=\pm \pi/3$ and $\pi$. 
At $\Theta=\pm \pi/3$ and $\pi$, thus, a mechanism similar to 
the Dashen phenomena at $\theta=\pi$ takes place.  
The transitions at $\Theta=\pm \pi/3$ and $\pi$ 
are called the RW transition. 
The  C breaking at $\Theta=\pm \pi/3$ and $\pi$ occurs 
when $T$ is high~\cite{Kouno}, while 
the P breaking at $\theta=\pi$ takes place when $T$ is zero~\cite{Witten} 
and then small. 
The grand canonical partition function at $\Theta =0$ is a sum of 
the canonical partition function over the quark number that is obtained by 
the Fourier transform of the grand canonical partition function with finite 
$\Theta$~\cite{RW,FK}. 
This means that the singular behavior of the RW transition in the vicinity of 
$\Theta =\pi/3$ 
reflects on the behavior of QCD at $\Theta =0$. 
Actually, it is confirmed in Ref.~\cite{Kouno} that the deconfinement 
crossover at $\Theta =0$ is a remnant of the first-order RW transition 
at $\Theta =\pi/3$. 
Thus, the thermodynamics at $\Theta =\pi/3$ is closely related to 
that at $\Theta =0$.
Furthermore, we can expect 
from the analogy between ${\cal L}_{\theta}$ and ${\cal L}_{\mu}$ that 
the thermodynamics at $\theta =\pi$ is also closely related to 
that at $\theta =0$.

LQCD has the sign problem at finite $\theta$, but not at finite $\Theta$. 
Therefore, we can test an effective model at $\theta=0$ and $\Theta \ge 0$ 
and apply the model to the case of 
$\theta >0$ and $\Theta \ge 0$. 
As a candidate of such effective models, 
we can consider the Nambu--Jona-Lasinio (NJL) 
model~\cite{NJ1,Klevansky,HK,Buballa,AY,Fujii,Osipov,Kashiwa,FIK,Boer,Boer2} 
and the Polyakov-loop extended Nambu--Jona-Lasinio (PNJL) model~\cite{Meisinger,Dumitru,Fukushima,Ghosh,Megias,Ratti,Ciminale,Rossner,Hansen,Sasaki,Schaefer,Costa,Kashiwa1,Fu,Abuki,Sakai,Sakai2,Kashiwa5,Kouno,Hell,Sakai3,Bhattacharyya,Mizher,Fukushima3,Matsumoto,Sasaki-T,Sakai5,Gatto,Gatto:2010pt}. 
The NJL model can describe the chiral-symmetry breaking and 
the Dashen mechanism~\cite{FIK,Boer,Boer2}, 
but not the confinement mechanism and the RW transition. 
The PNJL model can treat the deconfinement and the RW transition~\cite{Kashiwa1, Sakai,Kouno} as well as the chiral symmetry breaking 
and CME~\cite{Fukushima3}. 
However, the PNJL model has a weak correlation (entanglement) 
between the chiral and the deconfinement transition compared 
with LQCD~\cite{Sakai,Sasaki-T}. 
In order to solve this problem, we recently proposed a new version of 
the PNJL model, i.e. the entanglement-PNJL (EPNJL) model~\cite{Sakai5}, 
that has a four-quark vertex depending on the Polyakov loop. 
The EPNJL model can reproduce not only 
the strong correlation between the two transitions 
without~\cite{D'Elia-3, Sakai5} and with the strong magnet field~\cite{D'Elia5,Gatto:2010pt} 
but also the quark-mass dependence of the order of 
the RW endpoint~\cite{D'Elia-3,Sakai5}. 

The chiral and the $\mathbb{Z}_3$ symmetries are not exact symmetry 
for QCD with physical quark mass. 
Hence, the chiral and deconfinement transitions can not be defined exactly. 
Indeed, these transitions are approximately defined by 
the chiral condensate and the Polykov loop.  
In addition, the two transitions are crossover at $\theta=\Theta=0$. 
These situations often 
make it complicated the relation between the two transitions. 
At $\theta=\Theta=\pi$, in contrast, 
P and C are exact symmetry, so that their spontaneous breakings are 
clearly defined by P-odd and C-odd quantities respectively. 
Thus, the region of $\theta=\Theta=\pi$ is suitable to investigate 
the interplay between two kinds of transitions. 
Furthermore, we can expect that the thermodynamics 
at $\theta=\Theta=\pi$ is closely related to that at $\theta=\Theta=0$. 

In this paper, we analyze P and C violations at finite $\theta$ and 
$\Theta$ and the interplay between them, using the PNJL and EPNJL models. 
Particularly at $\theta=\Theta=\pi$, P and C are exact symmetries, so the 
analysis is mainly focused on the region. 
We also investigate 
the relation between the P and C breakings at $\theta=\Theta=\pi$ and 
the chiral and deconfinement transitions at $\theta=\Theta=0$. 
The P breaking at finite $\theta$ was already studied for the case of 
$T=0$ by the NJL model~\cite{FIK,Boer,Boer2}, so 
the present analysis is concentrated on the case of finite $T$, 
since the PNJL and the EPNJL model are reduced to the NJL model 
in the limit of $T=0$. 

This paper is organized as follows. 
In section II, the PNJL model is explained briefly. 
In section III, C and P violations are numerically 
investigated, particularly at $\Theta =\pi/3$ and $\theta =\pi$; note that 
$\Theta =\pi/3$ is identical with $\Theta =\pi$ because of the RW periodicity. 
We also explain the EPNJL model briefly in this section. 
Section IV is devoted to summary. 

\section{PNJL model}
\label{PNJL}

Pioneering work on the parity violation and its restoration in the framework 
of the NJL model was done by Fujihara, Inagaki and Kimura~\cite{FIK}. 
Boer and Boomsma studied on this issue extensively~\cite{Boer,Boer2}. 
Here, we extend their formalism based on the NJL model to that on 
the PNJL model. 
The two-flavor ($N_{f}=2$) PNJL Lagrangian with the $\theta$-dependent anomaly term 
is given as 
\begin{eqnarray}
{\cal L}  &=& {\bar q}(i \gamma_\nu D^\nu -m)q 
          - {\cal U}(\Phi [A],{\Phi} [A]^*,T)
\nonumber\\
          &+& G_1\sum_{a=0}^3\left[({\bar q}\tau_a q)^2 
               +({\bar q}i\gamma_5 \tau_a q)^2\right] \nonumber\\
          &+&8G_2\left[e^{i\theta}\det{\left(\bar{q}_{\rm R}q_{\rm L}\right)}
               +e^{-i\theta}\det{\left(\bar{q}_{\rm L}q_{\rm R}\right)}\right], \label{eq:E1}
\end{eqnarray}
where $q=(u,d)$ denotes the two-flavor quark field, 
$m$ does the current quark-mass matrix ${\rm diag}(m_u,m_d)$, 
$\tau_0$ is the $2\times 2$ unit matrix, 
$\tau_i (a=1,2,3)$ is the Pauli matrices 
and $D^\nu=\partial^\nu+iA^\nu-i\mu\delta^{\nu}_{0}$. 
The field $A^\nu$ is defined as $A^\nu=\delta^{\nu}_{0}gA^0_a{\lambda^a\over{2}}$ with the gauge field $A^\nu_a$, the Gell-Mann matrix $\lambda_a$ and the gauge coupling $g$. 
In the NJL sector, $G_1$ denotes the coupling constant of the scalar and pseudoscalar-type four-quark interaction, and 
$G_2$ is the coupling constant of the Kobayashi-Maskawa-'t Hooft determinant interaction~\cite{KMK,'t Hooft} the matrix indices of which run in the flavor space. 
The Polyakov potential ${\cal U}$, defined later in (\ref{eq:E13}), 
is a function of the Polyakov loop $\Phi$ and 
its Hermitian conjugate $\Phi^*$, 
\begin{align}
\Phi      = {1\over{N_{\rm c}}}{\rm Tr} L,~~~~
\Phi^{*}  = {1\over{N_{\rm c}}} {\rm Tr}L^\dag 
\end{align}
with
\begin{align}
L({\bf x}) = {\cal P} \exp\Bigl[
                {i\int^\beta_0 d \tau A_4({\bf x},\tau)}\Bigr],
\end{align}
where ${\cal P}$ is the path ordering and $A_4 = iA_0 $. 
In the chiral limit ($m_{\rm u}=m_{\rm d}=0$), the Lagrangian density has the exact $SU(N_f)_{\rm L} \times SU(N_f)_{\rm R}\times U(1)_{\rm v} \times SU(3)_{\rm c}$  symmetry. 
The $U(1)_{\rm A}$ symmetry is explicitly broken if $G_2\neq 0$. 
The temporal component of the gauge field is diagonal in the flavor space, because the color and the flavor space are completely separated out in the present case. 
In the Polyakov gauge, $L$ can be written in a diagonal form in the color space~\cite{Fukushima}: 
\begin{align}
L 
=  e^{i \beta (\phi_3 \lambda_3 + \phi_8 \lambda_8)}
= {\rm diag} (e^{i \beta \phi_a},e^{i \beta \phi_b},
e^{i \beta \phi_c} ),
\label{eq:E6}
\end{align}
where $\phi_a=\phi_3+\phi_8/\sqrt{3}$, $\phi_b=-\phi_3+\phi_8/\sqrt{3}$
and $\phi_c=-(\phi_a+\phi_b)=-2\phi_8/\sqrt{3}$. 
The Polyakov loop $\Phi$ is an exact order parameter of the spontaneous 
${\mathbb Z}_3$ symmetry breaking in the pure gauge theory.
Although the ${\mathbb Z}_3$ symmetry is not an exact one 
in the system with dynamical quarks, it may be a good indicator of 
the deconfinement phase transition. 
Therefore, we use $\Phi$ to define the deconfinement phase transition.

For simplicity, we assume below that $m_{\rm u}=m_{\rm d}=m_0$. 
Furthermore, to remove the $\theta$ dependence of the determinant interaction, we transform the quark field $q$ into the new field $q^\prime$ as 
\begin{eqnarray}
q_R=e^{i{\theta\over{4}}}q^\prime_R,~~~~~q_L=e^{-i{\theta\over{4}}}q^\prime_L. 
\label{UAtrans}
\end{eqnarray}
Under this ${\rm U}(1)_{\rm A}$ transformation, 
the quark and antiquark condensates are also transformed as 
\begin{eqnarray}
\sigma &\equiv &\bar{q}q
=\cos{\left({\theta\over{2}}\right)}\sigma^\prime
+\sin{\left({\theta\over{2}}\right)}\eta^\prime ,
\nonumber\\
\eta &\equiv &\bar{q}i\gamma_5q
=-\sin{\left({\theta\over{2}}\right)}\sigma^\prime
+\cos{\left({\theta\over{2}}\right)}\eta^\prime ,
\nonumber\\
a_i &\equiv &\bar{q}\tau_iq
=\cos{\left({\theta\over{2}}\right)}a_i^\prime
+\sin{\left({\theta\over{2}}\right)}\pi_i^\prime ,
\nonumber\\
\pi_i &\equiv &\bar{q}i\tau_i \gamma_5q
=-\sin{\left({\theta\over{2}}\right)}a_i^\prime
+\cos{\left({\theta\over{2}}\right)}\pi_i^\prime , 
\label{Econdensates}
\end{eqnarray}
where $\sigma^\prime$ is defined by the same form as $\sigma$ but $q$ is 
replaced by $q^\prime$; this is the case also for other condensates 
$\eta^\prime$, $a_i^\prime$ and $\pi_i^\prime$. 
The Lagrangian density is rewritten with the new field $q^\prime$ as 
\begin{eqnarray}
{\cal L}  
&=& {\bar q^\prime}(i \gamma_\nu D^\nu -m_{0+}-im_{0-}\gamma_5)q^\prime 
          - {\cal U}(\Phi [A],{\Phi} [A]^*,T) 
\nonumber\\
          &+& G_1\sum_{a=0}^3\left[({\bar q^\prime }\tau_aq^\prime )^2 
               +({\bar q^\prime}i\gamma_5 \tau_aq^\prime)^2\right] 
\nonumber\\
          &+& 8G_2\left[\det{\left(\bar{q^\prime }_{\rm R}q_{\rm L}^\prime 
               \right)}
          +\det{\left(\bar{q^\prime}_{\rm L}q_{\rm R}^\prime\right)}\right] ,
          \label{eq:E1rewritten}
\end{eqnarray}
where $m_{0+}=m_0\cos{\left({\theta\over{2}}\right)}$ and 
$m_{0-}=m_0\sin{\left({\theta\over{2}}\right)}$. 
Making the mean field approximation and performing 
the path integral over the quark field, 
one can obtain the thermodynamic potential $\Omega$ (per volume) 
for finite $T$ and $\mu$: 
\begin{align}
\Omega =& -2 \int \frac{d^3{\rm p}}{(2\pi)^3}
         \Bigl[ 3 \{E_+ ({\rm p})+E_-({\rm p})\} \nonumber\\
        & + \frac{1}{\beta}
           \ln~ [1 + 3\Phi e^{-\beta E_+^{-} }
           +3\Phi^{*}e^{-2\beta E_+^{-}}
           + e^{-3\beta E_+^{-}}]
         \nonumber\\
       & + \frac{1}{\beta}
           \ln~ [1 + 3\Phi e^{-\beta E_-^{-} }
           +3\Phi^{*}e^{-2\beta E_-^{-}}
           + e^{-3\beta E_-^{-}}]
         \nonumber\\ 
       & + \frac{1}{\beta}
           \ln~ [1 + 3\Phi^* e^{-\beta E_+^{+} }
           +3\Phi e^{-2\beta E_+^{+}}
           + e^{-3\beta E_+^{+}}]
         \nonumber\\ 
       & + \frac{1}{\beta}
           \ln~ [1 + 3\Phi^* e^{-\beta E_-^{+} }
           +3\Phi e^{-2\beta E_-^{+}}
           + e^{-3\beta E_-^{+}}]
         \nonumber\\ 
        & +U+{\cal U},  
\label{eq:E12} 
\end{align}
where $E_+^{\pm}=E_+({\bf p}) \pm \mu$ and 
$E_-^{\pm}=E_-({\bf p}) \pm \mu$ with 
\begin{eqnarray}
E_{\pm}&=&\sqrt{{\bf p}^2+C \pm 2\sqrt{D}}, 
\label{E12a}\\
C&=&M^2+N^2+A^2+P^2, 
\label{E12c}\\
D&=&A^2M^2+P^2N^2+2APMN\cos{\varphi}+A^2P^2\sin^2{\varphi}
\nonumber\\
&=&(M{\bf A}+N{\bf P})^2+({\bf A}\times{\bf P})^2\ge 0
\\
\label{E12e}
M&=&m_{0+}-2G_+\sigma^\prime =m_{0+}-2(G_1+G_2)\sigma^\prime ,
\label{E12M}\\
N&=&m_{0-}-2G_-\eta^\prime =m_{0-}-2(G_1-G_2)\eta^\prime , 
\label{E12N}\\
{\bf A}&=&(-2G_-a_1^\prime ,-2G_-a_2^\prime ,-2G_-a_3^\prime ), 
\label{E12Ameson}\\
{\bf P}&=&(-2G_+\pi_1^\prime ,-2G_+\pi_2^\prime ,-2G_+\pi_3^\prime ), 
\label{E12Pion}\\
A&=&\sqrt{{\bf A}\cdot{\bf A}},~~~P=\sqrt{{\bf P}\cdot{\bf P}},~~~{\bf A}\cdot{\bf P}=AP\cos{\varphi},
\label{E12absoluteangle}\\
U&=&G_+({\sigma^\prime}^2+{\pi_a^\prime}^2)+G_-({a_a^\prime}^2+{\eta^\prime}^2) .
\label{E12U}
\end{eqnarray}
In the right-hand side of \eqref{eq:E12}, only the first term diverges. 
The term is then regularized by the three-dimensional momentum
cutoff $\Lambda$~\cite{Fukushima,Ratti}. 
Following Ref.~\cite{Boer,Boer2}, we introduce $c$ as 
$G_1=(1-c)G$ and $G_2=cG$, 
where $0\leq c\leq 1$ and $G>0$. 
Hence, the NJL sector has four parameter of $m_0$, $\Lambda$, $G$ and $c$. 
We put $m_0=5.5$MeV. 
The parameters $\Lambda$ and $G$ are so chosen as to reproduce 
the pion decay constant $f_\pi =93$MeV and the pion mass $m_\pi =139$MeV 
at vacuum. 
The remaining parameter $c$ is a free parameter. 
Although the exact value of $c$ is unknown, 
it is known from the analysis of the $\eta$-$\eta^\prime$ splitting 
in the three flavor model that $c\sim 0.2$ is favorable~\cite{FBO}. 
The value $c=0.2$ has been also used in Refs.~\cite{Boer,Boer2}. 
Therefore, we adopt $c=0.2$ in this paper. 
(Note that the PNJL and the EPNJL model are reduced to the NJL model 
in the limit of $T=0$ as is mentioned above. ) 
For comparison, we will also show the result of the NJL model 
that has the same parameter set as in the PNJL model. 

The Polyakov potential ${\cal U}$ of Ref.~\cite{Rossner} is fitted 
to LQCD data in the pure gauge theory at finite $T$~\cite{Boyd,Kaczmarek}: 
\begin{align}
&{\cal U} = T^4 \Bigl[-\frac{a(T)}{2} {\Phi}^*\Phi\notag\\
      &~~~~~+ b(T)\ln(1 - 6{\Phi\Phi^*}  + 4(\Phi^3+{\Phi^*}^3)
            - 3(\Phi\Phi^*)^2 )\Bigr], \label{eq:E13}\\
&a(T)   = a_0 + a_1\Bigl(\frac{T_0}{T}\Bigr)
                 + a_2\Bigl(\frac{T_0}{T}\Bigr)^2,
 ~~~b(T)=b_3\Bigl(\frac{T_0}{T}\Bigr)^3  \label{eq:E14} .
\end{align}
The parameters included in ${\cal U}$ are summarized in Table I.  
The Polyakov potential yields a first-order deconfinement phase transition at 
$T=T_0$ in the pure gauge theory. 
The original value of $T_0$ is $270$ MeV evaluated by the pure gauge lattice QCD calculation. 
However, the PNJL model with this value of $T_0$ yields somewhat larger value of the transition temperature at zero chemical potential than the full LQCD simulation~\cite{Karsch3,Karsch4,Kaczmarek2} predicts. 
Therefore, we rescale $T_0$ to 212~MeV in the numerical calculations~\cite{Schaefer}. 

\begin{table}[h]
\begin{center}
\begin{tabular}{llllll}
\hline
~~~~~$a_0$~~~~~&~~~~~$a_1$~~~~~&~~~~~$a_2$~~~~~&~~~~~$b_3$~~~~~
\\
\hline
~~~~3.51 &~~~~-2.47 &~~~~15.2 &~~~~-1.75\\
\hline
\end{tabular}
\caption{
Summary of the parameter set in the Polyakov-potential sector
used in Ref.~\cite{Rossner}. 
All parameters are dimensionless. 
}
\end{center}
\end{table}

The variables $X=\Phi$, ${\Phi}^*$ and $\sigma$ 
satisfy the stationary conditions, 
\begin{eqnarray}
\partial \Omega/\partial X=0. 
\label{eq:SC}
\label{condition}
\end{eqnarray}
The solutions of the stationary conditions do not give 
the global minimum $\Omega$ 
necessarily. 
There is a possibility that they yield a local minimum or even a maximum. 
We then have checked that the solutions yield the global minimum when the solutions $X(T,\theta ,\Theta )$ are inserted into (\ref{eq:E12}).

Now we consider the imaginary chemical potential $\mu =i\Theta T$. 
The thermodynamic potential $\Omega$ of \eqref{eq:E12} 
has a periodicity of $2\pi$ in 
both $\theta$ and $\Theta$. Hereafter, we mainly consider one circle, 
$0\leq \theta \le 2\pi$ and $0 \le \Theta \le 2\pi$. 
In addition, $\Omega$ has the RW periodicity:  
\begin{eqnarray}
\Omega(T,\theta ,\Theta )=\Omega (T,\theta ,\Theta +{2\pi\over{3}})=\Omega (T,\theta ,\Theta +{4\pi\over{3}})
\label{RW-periodicity}
\end{eqnarray}
for $-2\pi/3<\Theta\le 0$. This is understood as follows. 
The thermodynamical potential $\Omega$ is not 
invariant under the ${\mathbb Z}_3$ transformation, 
\begin{eqnarray}
\Phi  \to \Phi e^{-i{2\pi k/{3}}} \;,\quad
\Phi ^{*} \to \Phi^{*} e^{i{2\pi k/{3}}} \;, 
\end{eqnarray}
while ${\cal U}$ of (\ref{eq:E13}) is invariant. 
Instead of the ${\mathbb Z}_3$ symmetry, however, $\Omega$ is invariant 
under the extended ${\mathbb Z}_3$ transformation~\cite{Sakai}, 
\begin{align}
&e^{\pm i \Theta} \to e^{\pm i \Theta} e^{\pm i{2\pi k\over{3}}},\quad  
\Phi \to \Phi e^{-i{2\pi k\over{3}}}, 
\notag\\
&\Phi^{*} \to \Phi^{*} e^{i{2\pi k\over{3}}} .
\label{eq:K2}
\end{align}
This invariance means that $\Omega$ has the Roberge-Weiss 
periodicity~\cite{Sakai}. 
This can be seen more explicitly by introducing 
the modified Polyakov loop 
\begin{eqnarray}
\Psi \equiv \Phi e^{i\Theta}
\label{Psi}
\end{eqnarray}
invariant under the extended ${\mathbb Z}_3$ 
transformation (\ref{eq:K2}). 
The thermodynamic potential $\Omega$ is then rewritten with 
the modified Polykov loop as 
\begin{align}
\Omega =& -2 \int \frac{d^3{\rm p}}{(2\pi)^3}
         \Bigl[ 3 \{E_+ ({\rm p})+E_-({\rm p})\} \nonumber\\
        & + \frac{1}{\beta}
           \ln~ [1 + 3\Psi e^{-\beta E_+({\bf p})}
         \nonumber\\
&~~~~~~~~~~~~~~~~
+3\Psi^{*}e^{-2\beta E_+({\bf p})+i3\Theta}
+ e^{-3\beta E_+({\bf p})+3i\Theta }]
         \nonumber\\
        & + \frac{1}{\beta}
           \ln~ [1 + 3\Psi e^{-\beta E_-({\bf p})}
         \nonumber\\
&~~~~~~~~~~~~~~~~
+3\Psi^{*}e^{-2\beta E_-({\bf p})-3i\Theta}
+ e^{-3\beta E_-({\bf p})-3i\Theta }]
         \nonumber\\ 
        & + \frac{1}{\beta}
         \ln~ [1 + 3\Psi^* e^{-\beta E_+({\bf p})}
         \nonumber\\
&~~~~~~~~~~~~~~~~
+3\Psi e^{-2\beta E_+({\bf p})+3i\Theta }
+ e^{-3\beta E_+({\bf p})+3i\Theta }]
         \nonumber\\
        & + \frac{1}{\beta}
           \ln~ [1 + 3\Psi^* e^{-\beta E_-({\bf p})}
         \nonumber\\
&~~~~~~~~~~~~~~~~
+3\Psi e^{-2\beta E_-({\bf p})+3i\Theta }
+ e^{-3\beta E_-({\bf p})+3i\Theta}]
         \Bigl]\nonumber\\
        & +G_+({\sigma^\prime}^2+{\pi_a^\prime}^2)+G_-({a_a^\prime}^2+{\eta^\prime}^2) \nonumber\\
        & +T^4 \Bigl[-\frac{a(T)}{2} {\Psi}^*\Psi 
        + b(T)\ln(1 - 6{\Psi\Psi^*}  
\nonumber\\
        &~~~~~~~~~~~+ 4(\Psi^3e^{-3i\Theta}+{\Psi^*}^3e^{3i\Theta})
            - 3(\Psi\Psi^*)^2 )\Bigr]. 
\label{reOmega}
\end{align}
In \eqref{reOmega}, $\Omega$ depends on $\Theta$ only 
through $e^{i3\Theta}$. Thus, $\Omega$ has the RW periodicity 
\eqref{RW-periodicity}.  

The thermodynamic potential $\Omega$ is invariant under P transformation, 
\bea
\eta \to -\eta,~~~~\pi_a \to -\pi_a. 
\eea
for $\theta=0$ and $\pi$. 
The potential $\Omega$ is also invariant under C transformation, 
\bea
\pi_2 \rightarrow -\pi_2,~~~~a_2 \rightarrow -a_2,
~~~~\Phi \leftrightarrow \Phi^* ,
\eea
for $\Theta=0$ and $\pi$. 
Because of the RW periodicity \eqref{RW-periodicity}, the C-invariance is true also for $\Theta=n\pi/3$, where $n$ is an arbitrary integer. 
Hereafter, we mainly consider a period $0\le \Theta \le \pi/3$ for simplicity. 

For $\Theta=\pi/3$, C-symmetry is 
spontaneously broken at $T$ higher than 
some critical temperature $T_{\rm C}$~\cite{Kouno}. 
C-odd quantities such as the phase of $\Psi$, the imaginary part of $\Psi$ 
or the baryon number density is the order parameter of this phase transition. 
For $\theta=\pi$, as shown later, P-symmetry is spontaneously broken when 
$c$ is greater than some critical value $c_{\rm cri}$ and $T$ is 
smaller than some critical temperature $T_{\rm P}$~\cite{FIK,Boer,Boer2}. 
P-odd quantities such as $\eta$ are the order parameter 
of this phase transition.

\section{P and C breaking at finite $\theta$ and $\Theta$}
\label{theta}

In this section, P and C violations are investigated first by the PNJL model. 
After the PNJL analysis, we use the EPNJL model to explain 
the strong correlation between the chiral and deconfinement transitions at 
zero and finite $\theta$ and $\Theta$.

The present PNJL model has eight condensates of quark-antiquark pair. 
However, $\vec{a}$ and $\vec{\pi}$ vanish~\cite{Boer,Boer2}, 
since $m_{\rm u}=m_{\rm d}$ and the isospin chemical potential 
is not considered here. 
We can then concentrate ourselves on $\sigma$, $\eta$ and $\Phi$.

\subsection{The case of $\theta =\Theta =0$}
\label{zero}
\begin{figure}[htbp]
\begin{center}
 \includegraphics[width=0.35\textwidth]{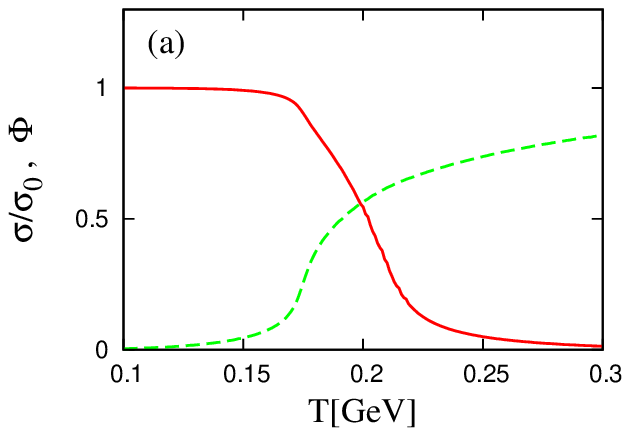}
 \includegraphics[width=0.35\textwidth]{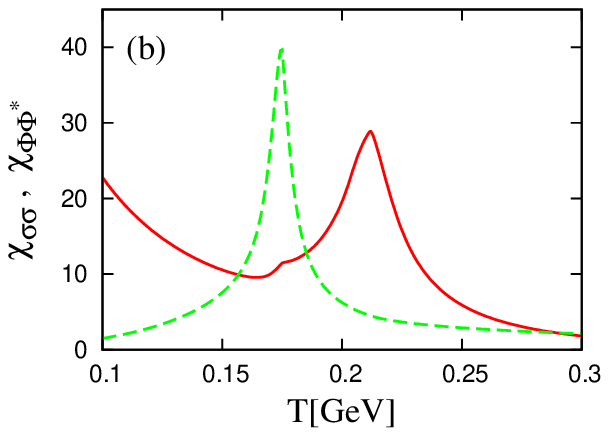}
 \end{center}
\caption{(color online). $T$ dependence of (a) the chiral condensate and 
the Polyakov loop and also (b) their susceptibilities 
at $\theta=0$ and $\Theta =0$. 
The solid (dashed) line represents 
the chiral condensate (Polyakov loop) in (a) 
and the chiral (Polyakov-loop) susceptibility in (b). 
In panel (b), $\chi_{\Phi\Phi^*}$ is multiplied by $10^2$. }
\label{fig-zero}
\end{figure}

In this subsection, we consider the case of $\theta =\Theta =0$. 
Figure~\ref{fig-zero}(a) shows $T$ dependence of the chiral condensate 
$\sigma$ and the Polyakov loop $\Phi$; here, 
$\sigma$ is normalized by the value $\sigma_0$ at $T=0$. 
Rapid but smooth changes in $\sigma$ (solid line) and $\Phi$ (dashed line) 
indicate that the chiral-symmetry restoration and the deconfinement 
transition are crossover. 
Figure~\ref{fig-zero}(b) presents the chiral and Polyakov-loop 
susceptibilities, 
$\chi_{\sigma\sigma}$ and $\chi_{\Phi\Phi^*}$, as a function of $T$. 
The pseudocritical temperatures, $T_{\chi}$ and $T_{\rm d}$, of 
the chiral-symmetry restoration and the deconfinement transition
are defined by peaks of $\chi_{\sigma\sigma}$ (solid line) 
and $\chi_{\Phi\Phi^*}$ (dashed line); 
here, the susceptibilities are normalized by $T$ so as 
to be dimensionless~\cite{Kashiwa1}. 
In the present PNJL model, $T_{\chi}=216$~MeV and $T_{\rm d}=173$~MeV, 
while $T_{\chi} \approx T_{\rm d} \approx 173$~MeV 
in LQCD~\cite{Karsch3,Karsch4,Kaczmarek}. 
The value of $T_{\chi}$ is even larger than 
the NJL result $T_{\chi}^{\rm NJL}=186$~MeV. 
Thus, the present PNJL result is consistent 
with LQCD data for $T_{\rm d}$, but not for $T_{\chi}$. 
This will be discussed at the end of this section.

\subsection{The case of $\theta \ge 0$ and $\Theta=0$}
\label{finite-theta}

\begin{figure}[htbp]
\begin{center}
 \includegraphics[width=0.35\textwidth]{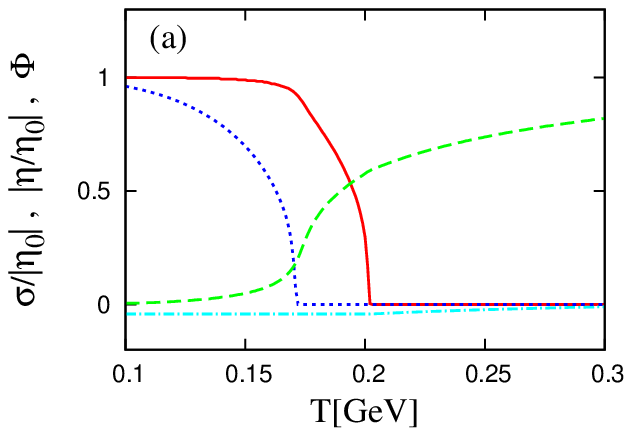}
 \includegraphics[width=0.45\textwidth]{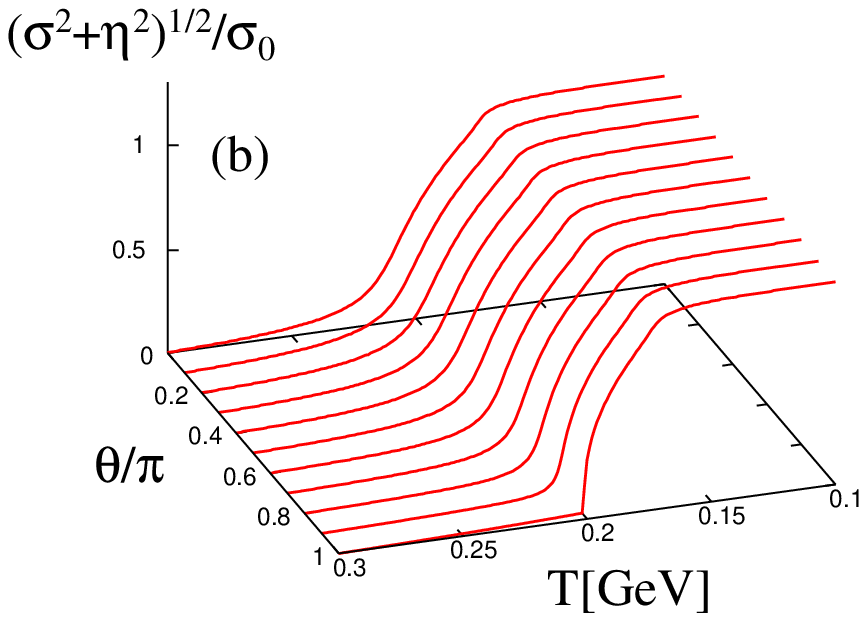}
 \end{center}
\caption{(color online). (a) $T$ dependence of $\sigma$, $\Phi$ and 
the absolute value of $\eta$ at $\theta =\pi$ 
and also (b) $\sqrt{\sigma^2+\eta^2}$ in $T$-$\theta$ plane. 
In panel (a), 
the solid (dashed) line represents 
$|\eta / \eta_0|$ ($\Phi$), while 
the dot-dashed line stands for $\sigma/|\eta_0|$. 
The dotted line shows $|\eta/\eta_0|$ calculated with the NJL model.  
In panel (b), the solid lines show the PNJL results at several values of 
$\theta$. In both panels, $\Theta =0$. 
}
\label{fig-finite-th}
\end{figure}

In this subsection, $\Theta$ is fixed at zero. 
Figure~\ref{fig-finite-th}(a) shows $T$ dependence of $\sigma$, $\Phi$ and 
the absolute value of $\eta$ at $\theta =\pi$; 
here, $\sigma$ and $|\eta|$ are normalized by $|\eta_0|$ 
the value of $|\eta|$ at $T=0$. 
The solid and dashed lines represent 
$|\eta /\eta_0|$ and $\Phi$, respectively, while 
the dot-dashed line stands for $\sigma/|\eta_0|$. 
The order parameter $\eta$ of P violation is finite 
at $T < T_{\rm P}=202$~MeV, while zero at $T > T_{\rm P}$. 
This means that P-symmetry is spontaneously broken below $T_{\rm P}$ and 
restored above $T_{\rm P}$. 
The restoration of P breaking at $T = T_{\rm P}$ is a second-order transition, 
as shown by the solid line. 
$T$ dependence of $\Phi$ little changes between Fig.~\ref{fig-zero}(a)
and Fig.~\ref{fig-finite-th}(a), 
indicating that $T_{\rm d}=173$~MeV also for $\theta =\pi$. 
Thus, $T_{\rm P}$ is much higher than $T_{\rm d}$. 
Comparing the solid and the dot-dashed lines shows that 
$|\sigma|$ is much smaller than $|\eta|$ at small $T$. Thus, 
the thermodynamics at $\theta =\pi$ is mainly controlled by 
$|\eta|$. 

The previous works on the NJL model showed for $\theta =\pi$ that 
P symmetry can be violated for 
low $T$, but not for high $T$~\cite{FIK,Boer,Boer2}. 
This statement is supported by the present PNJL model. 
In Fig.~\ref{fig-finite-th}(a), the dotted line is a result of the NJL model 
for $|\eta /\eta_0|$. The order of P restoration is 
second order~\cite{FIK,Boer,Boer2}, as shown by the dotted line. 
Thus, the PNJL and the NJL model show that the P restoration  is 
a second-order transition, while the linear sigma model 
points out that it is a first-order transition~\cite{Mizher2,Boer2}. 
In the present NJL model, $T_{\rm P} \approx 172$~MeV and then close 
to $T_{\rm d}$, but this is just accidental. 

Figure~\ref{fig-finite-th}(b) shows $T$ dependence of 
$\Sigma\equiv \sqrt{\sigma^2+\eta^2}$ at several values of $\theta$. 
For $\theta=0$, $\Sigma$ agrees with $|\sigma|$ the approximate order 
parameter of the chiral symmetry. For $\theta=\pi$, meanwhile, 
$\Sigma$ is close to the order parameter $|\eta|$ of P violation. 
Although $\Sigma$ itself is not an order parameter of P restoration, 
it has a cusp at $T=T_{\rm P}$ as a reflection of the second-order 
P restoration at $T=T_{\rm P}$. 
This nonanalytic behavior in $\Sigma$ is smeared out as $\theta$ decreases 
from $\pi$, but a smooth but rapid $T$ dependence remains in $\Sigma$ 
at $\theta =0$. 
Thus, the crossover chiral symmetry restoration 
at $\theta =0$ can be regarded as a remnant of the second-order P restoration 
at $\theta =\pi$. 

\begin{figure}[htbp]
\begin{center}
  \includegraphics[width=0.35\textwidth]{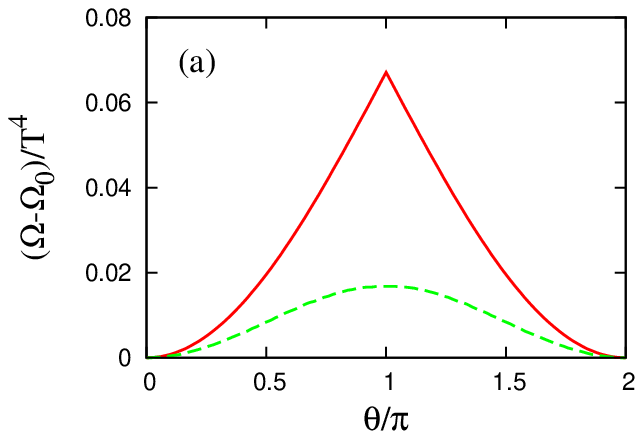}
  \includegraphics[width=0.35\textwidth]{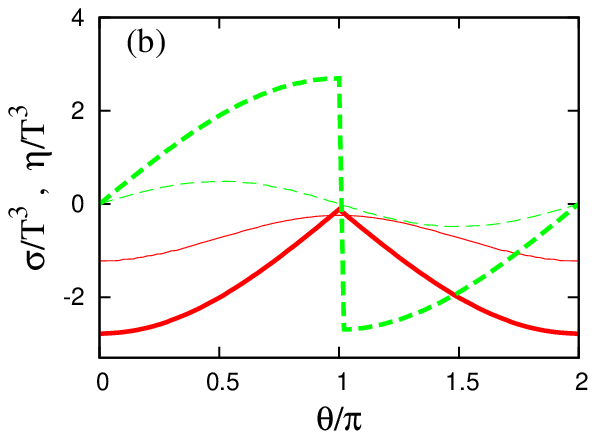}
  \includegraphics[width=0.35\textwidth]{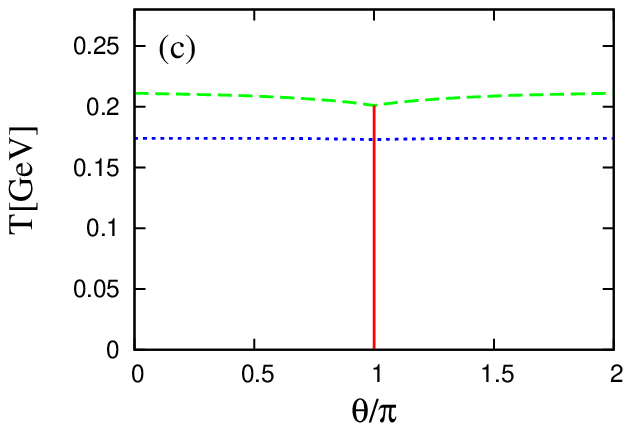}
 \end{center}
\caption{(color online). 
(a) $\theta$ dependence of the thermodynamic potential. 
The dashed (solid) line represents the result at $T=220$MeV (160MeV). 
$\Omega_0\equiv \Omega (\theta =0, \Theta =0, T)$. 
(b) $\theta$ dependence of $\sigma$ and $\eta$, 
The solid (bold solid) line represents $\sigma$ at $T=220$MeV (160MeV), 
while the dashed (bold dashed) line represents the $\eta$ at $T=220$MeV (160MeV). 
(c) Phase diagram in $T$-$\theta$ plane at $\Theta=0$. 
The vertical solid line represents the first-order phase transition caused by P violation. 
The dashed and dotted lines stand for 
the chiral and the deconfinement crossover, respectively. 
}
\label{fig-T-th}
\end{figure}

Figure~\ref{fig-T-th}(a) show $\theta$ dependence of the thermodynamical potential $\Omega$. 
We see that $\Omega$ is minimum at $\theta =0$ and is maximum at $\theta =\pi$. 
When $T<T_{\rm P}$ (solid line), due to the P violation at $\theta =\pi$, $\Omega$ has a cusp there. 

Figure~\ref{fig-T-th}(b) show $\theta$ dependence of $\sigma$ (solid and bold solid lines) and $\eta$ (dashed and bold dashed lines). 
We see that, at low temperature, due to the P violation, $\sigma$ has a cusp at $\theta =\pi$, while $\eta$ is discontinuous there. 
In general, below the critical temperature $T_{\rm P}$, $\theta$-even quantities such as $\Omega$ or $\sigma$ have a cusp at $\theta =\pi$, while $\theta$-odd quantities such as $\eta$ are discontinuous there. 

Figure~\ref{fig-T-th}(c) shows the phase diagram in $T$-$\theta$ plane at 
$\Theta=0$. 
The vertical solid line represents the first-order phase transition 
induced by spontaneous P breaking, since on the line $\Omega$ is not smooth 
in the $\theta$ direction. 
The dashed line stands for the chiral crossover connected with 
the solid line at the endpoint of P violation. The order of the endpoint is 
the second order, as mentioned above. The dotted line is the 
deconfinement-crossover line that is almost independent of $\theta$. 

\subsection{The case of $\Theta \ge 0$ and $\theta=0$}
\label{finite-imaginary}

\begin{figure}[htbp]
\begin{center}
 \includegraphics[width=0.35\textwidth]{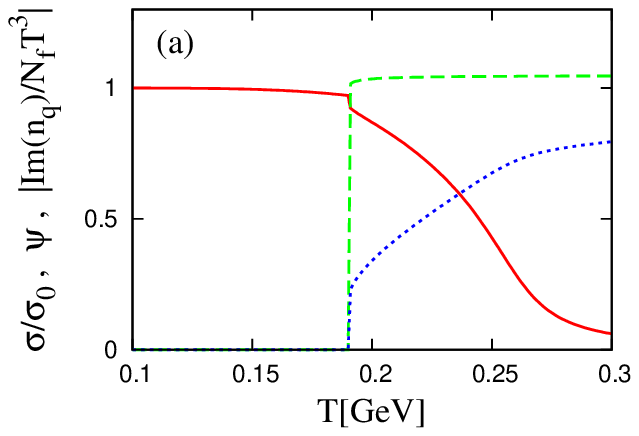}
 \includegraphics[width=0.45\textwidth]{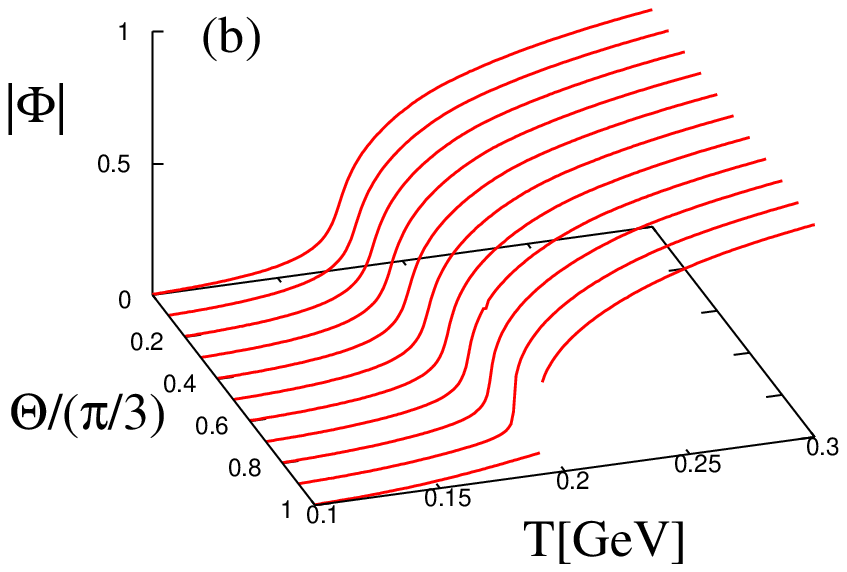}
\end{center}
\caption{(color online). (a) $T$ dependence of $\sigma$ (solid line), the absolute value of the phase $\psi$ of $\Psi$ (dashed line) and the absolute value of the imaginary part of the quark nunber density $n_q$ divided by $N_fT^3$ (dotted line) 
at $\Theta =\pi/3$ and also (b) the absolute value 
of $\Phi$ in $T$-$\Theta$ plane. 
In both cases, $\theta =0$. }
\label{fig-finite-imaginary}
\end{figure}

In this subsection, we review the case of finite $\Theta$ to see the analogy 
between P violation induced by finite $\theta$ and C violation 
induced by finite $\Theta$. 
For this purpose, $\theta$ is fixed at 0 here. 
Figure \ref{fig-finite-imaginary}(a) shows $T$ dependence of 
$\sigma$, the absolute value of the phase $\psi$ of 
$\Psi$, the imaginary part of the quark number density $n_q$ at $\Theta =\pi /3$. 
As shown by the dashed (dotted) line, 
$\psi$ (${\rm Im}(n_q)$) is zero below $T_{\rm C}=189$~MeV but finite above $T_{\rm C}$. 
This indicates that C symmetry is spontaneously broken 
above $T_{\rm C}$~\cite{Kouno}. 
The order of the C violation (RW transition) is of first order, 
since $\Omega$ is not smooth in the $\Theta$ direction~\cite{Sakai}. 
The order of the endpoint at $T=T_{\rm C}$ 
of the C violation  line 
depends on the Polyakov potential ${\cal U}$ used. 
It is the second order~\cite{Kouno} 
for ${\cal U}$ of Ref.~\cite{Fukushima}, but 
the first order~\cite{Sakai3} for ${\cal U}$ of Ref.~\cite{Rossner}.
The agreement of the PNJL result with the LQCD data at finite $\Theta$ is 
better in the later  than in the former~\cite{Sakai3}. Therefore, we take 
the latter case in this paper. Hence, the endpoint of the C violation 
(RW transition) is {\it a triple point} in the present case; 
this is explicitly shown in Ref.~\cite{Sakai3}. 
Because of this first-order phase transition, $\sigma$ (solid line) 
has a gap at $T=T_{\rm C}$. 
This behavior of $\sigma$ is easily understood by the discontinuity theorem on 
the first-order phase transition 
by Barducci, Casalbuoni, Pettini and Gatto~\cite{BCPG}. 
However, the absolute value of $\sigma$ is still large at $T=T_{\rm C}$, 
as shown by the solid line. 
Hence the chiral transition should be regarded as not of first order but crossover. 
Comparing Fig.~\ref{fig-finite-imaginary}(a) with Fig.~\ref{fig-zero}(a), we can see that the chiral crossover restoration is slower at $\Theta =\pi/3$ than at $\Theta =0$.

Figure~\ref{fig-finite-imaginary}(b) shows the absolute value of the Polyakov loop $\Phi$ in $T$-$\Theta$ plane; note that $|\Psi |=|\Phi |$ by definition. 
The absolute value of $\Phi$ is not an order parameter of 
C violation, but $|\Phi|$ (solid line) 
has a jump near $T=T_{\rm C}$ and $\Theta=\pi/3$ as a reflection of 
the fact that the endpoint of C violation is of first order 
at $T=T_{\rm C}$ and $\Theta=\pi/3$. 
This nonanalytic behavior in $|\Phi |$ is smeared out as $\Theta$ decreases 
from $\pi/3$. However, a smooth but rapid $T$ dependence remains in $|\Phi|$. 
Thus, the deconfinement crossover at $\Theta =0$ is a remnant 
of C violation at $\Theta =\pi /3$~\cite{Kouno}. 

\begin{figure}[htbp]
\begin{center}
 \includegraphics[width=0.35\textwidth]{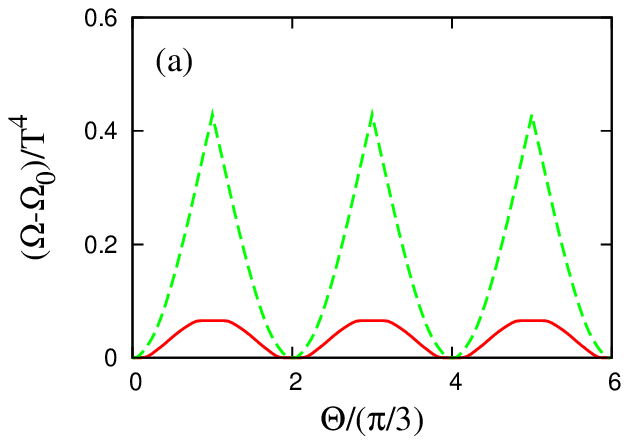}
 \includegraphics[width=0.35\textwidth]{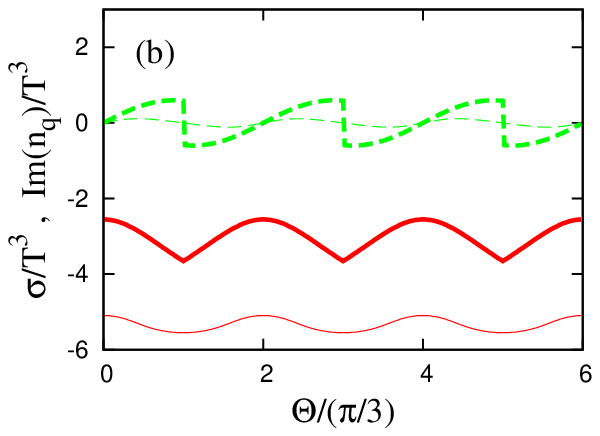}
 \includegraphics[width=0.35\textwidth]{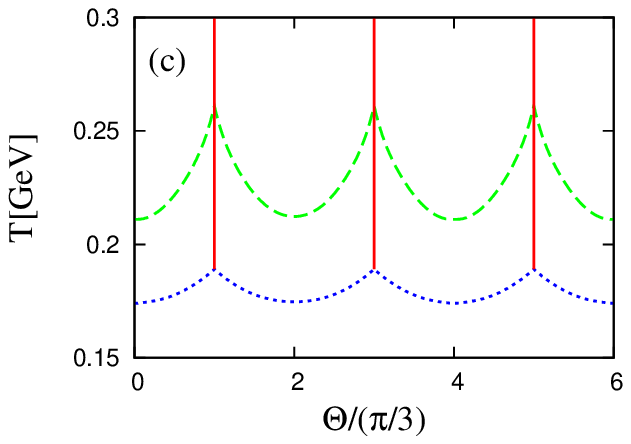}
\end{center}
\caption{(color online). 
(a) $\Theta$ dependence of thermodynamic potential at $\theta =0$. 
The solid (dashed) line represents the result for $T=$175(195)MeV. 
$\Omega_0\equiv \Omega (\theta =0, \Theta =0, T)$. 
(b) $\Theta$ dependence of $\sigma$ and ${\rm Im}(n_{q})$ at $\theta =0$. 
The solid (bold solid) line represents $\sigma$ at $T=175(195)$MeV, while the dashed (bold dashed) line represents ${\rm Im}(n_{q})$ at $T=175(195)$MeV. 
(c)The phase diagram in $T$-$\Theta$ plane at $\theta =0$. 
The vertical solid line represents the first-order phase transition 
of C violation (the RW transition). 
The dashed line stands for the chiral crossover, while 
the dotted line shows the deconfinement crossover. 
}
\label{fig-finite-T-imaginary}
\end{figure}

Figure~\ref{fig-finite-T-imaginary}(a) show the $\Theta$ dependence of the thermodynamical potential $\Omega$. 
The RW periodicity (\ref{RW-periodicity}) is clearly seen in this figure. 
We also see that $\Omega$ is minimum at $\Theta =0$ ($2\pi /3$, $4\pi /3$ ) and is maximum value at $\Theta =\pi /3$ ($\pi$, $5\pi /3$). 
When $T>T_{\rm C}$ (dashed line), due to the C violation at $\Theta =\pi /3$ ($\pi$, $5\pi /3$), $\Omega$ has a cusp there. 

Figure~\ref{fig-finite-T-imaginary}(b) show the $\Theta$ dependence of $\sigma$ (solid and bold solid lines) and the imaginary part of the quark number density $n_{q}$ (dashed and bold dashed lines). 
We see that, at high temperature, due to the C violation, $\sigma$ has a cusp at $\Theta =\pi /3$ ($\pi$, $5\pi /3$), while ${\rm Im}(n_q)$ is discontinuous there. 
In general, above the critical temperature $T_{\rm C}$, $\Theta$-even quantities such as $\Omega$ or $\sigma$ have a cusp at $\Theta =\pi /3$ ($\pi$, $5\pi /3$), while $\Theta$-odd quantities such as ${\rm Im}(n_q)$ are discontinuous there. 

The phase diagram in the $T$-$\Theta$ plane is shown 
in Fig.~\ref{fig-finite-T-imaginary}(c). 
The vertical solid line represents the first-order RW transition induced by 
spontaneous C-breaking, since on the line $\Omega$ is not smooth 
in the $\Theta$ direction. 
The dashed line stands for the chiral crossover. 
The dotted line is the deconfinement-crossover line connected with 
the solid line at the endpoint of C violation. The order of the endpoint is 
of first order, as mentioned above. 

\subsection{The case of $\theta \ge 0$ and $\Theta \ge 0$}
\label{finite-th-imaginary}

\begin{figure}[htbp]
\begin{center}
 \includegraphics[width=0.35\textwidth]{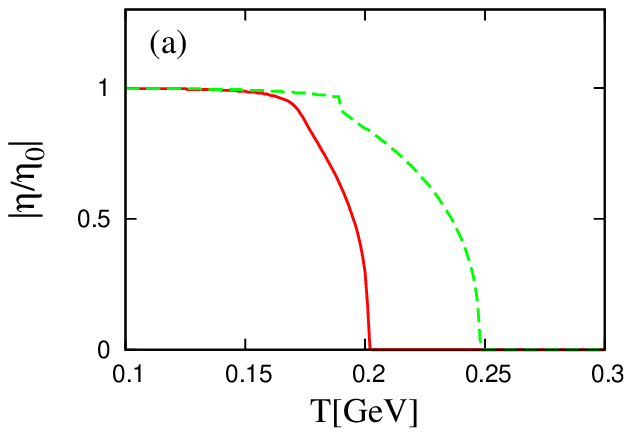}
 \includegraphics[width=0.35\textwidth]{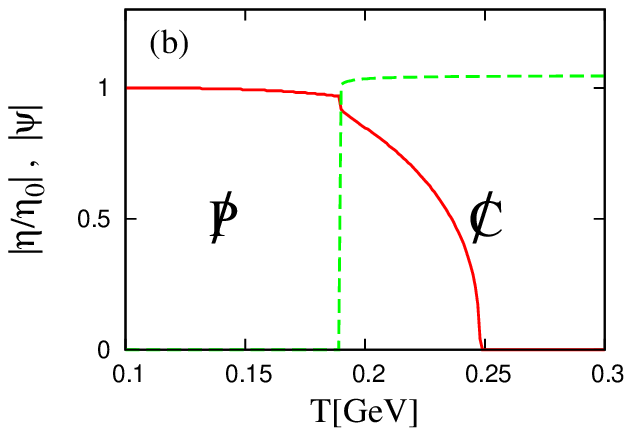}
 \end{center}
\caption{ (color online). $T$ dependence of (a) 
$|\eta|$ at $\Theta =0$ and $\Theta =\pi /3$ and also 
(b) $|\eta /\eta_0|$ and $|\psi|$ at $\Theta =\pi /3$. 
Here, $\theta$ is fixed at $\pi$. 
In panel (a), the solid (dashed) line represents the PNJL result 
at $\Theta =0$ ($\pi/3$). In panel (b), the solid (dashed) line stands for 
$\eta $ ($\psi$). 
}
\label{fig-finite-th-imaginary}
\end{figure}

In this subsection, we consider the case of finite $\theta$ and $\Theta$ to 
examine the correlation between C violation and P restoration. 
Figure~\ref{fig-finite-th-imaginary}(a) shows 
$T$ dependence of $|\eta|$ at $\Theta =0$ and $\pi /3$ in the case of 
$\theta=\pi$. 
The solid and the dashed line show the PNJL results 
at $\Theta =0$ and $\pi /3$, respectively.  
The critical temperature $T_{\rm P}$ of P-restoration is 202~MeV for 
$\Theta =0$ and 249~MeV for $\Theta =\pi/3$. 
Thus, the imaginary chemical potential enhances the P violation as well as 
the chiral symmetry breaking. 
Figure~\ref{fig-finite-th-imaginary}(b) shows $T$ dependence of $|\eta|$ and 
$|\psi|$ at $\Theta =\pi/3$ and $\theta =\pi$; here, $|\eta|$ is 
normalized by the value $|\eta_0|$ at $T=0$. 
The solid and the dashed line stand for $|\eta|$ and $|\psi|$, respectively. 
A first-order C violation occurs at $T_{\rm C}=190$~MeV, 
while a second-order P-restoration takes place at $T_{\rm P}=249$~MeV. 
Thus, $T_{\rm P}$ is much larger than $T_{\rm C}$. 
This indicates that either P or C symmetry or both are broken for any $T$. 
The inequality $T_{\rm C} \ll T_{\rm P}$ at $\theta =\pi$ and $\Theta =\pi /3$ shown in Fig.~\ref{fig-finite-th-imaginary}(b) and the inequality 
$T_{\rm d} \ll T_{\chi}$ at $\theta =0$ and $\Theta =0$ 
shown in Fig.~\ref{fig-zero}(b) come from the fact that 
the correlation between the chiral and deconfinement transitions are weak in 
the present PNJL model. This problem will be solved in the next subsection 
by using the EPNJL model. 

\begin{figure}[htbp]
 \begin{center}
  \includegraphics[width=0.35\textwidth]{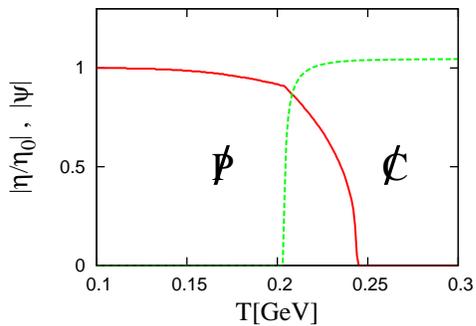}
 \end{center}
\caption{ (color online). $T$ dependence of $|\eta /\eta_0|$ (solid line) and $|\psi |$ (dashed line) at $\theta =\pi$ and $\Theta =\pi /3$ in the PNJL model with the Polyakov potential proposed by Fukushima\cite{Fukushima}. }
\label{fig_F_PC}
\end{figure}

The order of C violation depends on 
the Polyakov potential $\cal {U}$ taken. 
As mentioned above, the order is of first order for 
$\cal {U}$ of \eqref{eq:E13} proposed by Ro\"{s}ner et al.~\cite{Rossner}. 
Meanwhile, as shown in Fig.~\ref{fig_F_PC}, it is of second order 
for $\cal {U}$ proposed by Fukushima~\cite{Fukushima},  
\begin{align}
&{\cal U} = -bT\Bigl[54e^{-a/T}{\Phi}^*\Phi\notag\\
      &~~~~~+ \ln(1 - 6{\Phi\Phi^*}  + 4(\Phi^3+{\Phi^*}^3)
            - 3(\Phi\Phi^*)^2 )\Bigr] ,  \label{eq:add1}
\end{align}
where $a=664$MeV and $b=0.015\Lambda^3$~\cite{Kouno}. 
However, the first order is more plausible, since 
$\cal U$ proposed by Ro\"{s}ner et al. is 
more consistent with LQCD data at imaginary chemical potential 
than that by Fukushima~\cite{Sakai3}.  
Furthermore, 8-flavor LQCD data 
at imaginary isospion chemical potential~\cite{Cea} 
favors $\cal U$ proposed by Ro\"{s}ner et al, since 
the EPNJL model with $\cal U$ proposed by Rossener et al. 
yields results consistent with the LQCD data~\cite{Sakai5}.

\subsection{The EPNJL model}
\label{EPNJL}

Since the four-quark coupling constant $G$ contains effects of gluons, 
$G$ may depend on $\Phi$. 
In fact, recent calculations~\cite{Braun,Kondo,Herbst} of 
the exact renormalization group equation (ERGE)~\cite{Wetterich} suggest 
that the chiral and the deconfinement transitions coincide 
by the higher-order mixing interaction induced by ERGE. 
It is highly expected that the functional form and the strength of 
the entanglement vertex are determined in future by these theoretical 
approaches. In Ref.~\cite{Sakai5}, we assumed 
the following $\Phi$ dependence of $G$ by respecting the chiral symmetry, 
P symmetry, C symmetry and the extended $\mathbb{Z}_3$ symmetry, 
\begin{eqnarray}
G(\Phi)=G[1-\alpha_1\Phi\Phi^*-\alpha_2(\Phi^3+\Phi^{*3})]. 
\label{entanglement-vertex}
\end{eqnarray}
This model is called the entanglement PNJL (EPNJL). 
The EPNJL model with the parameter set, 
$\alpha_1=\alpha_2=0.2$ and $T_0=190$~MeV, can reproduce 
LQCD data at imaginary chemical potential and 
real isospin chemical potential 
as well as the results at zero chemical potential~\cite{Sakai5}. 
The EPNJL model with this parameter set is applied for the present case 
with zero and finite $\theta$ and $\Theta$. 

\begin{figure}[htbp]
\begin{center}
 \includegraphics[width=0.35\textwidth]{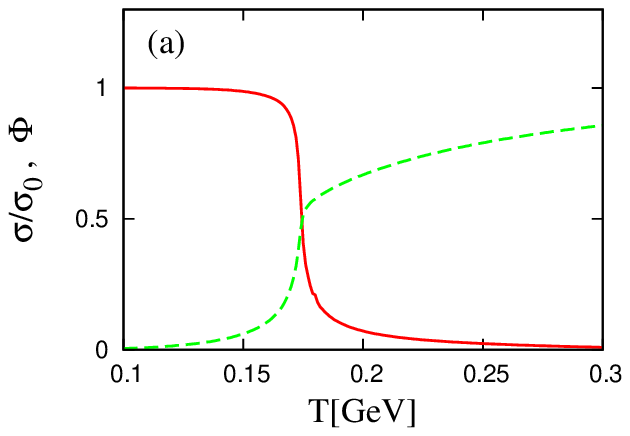}
 \includegraphics[width=0.35\textwidth]{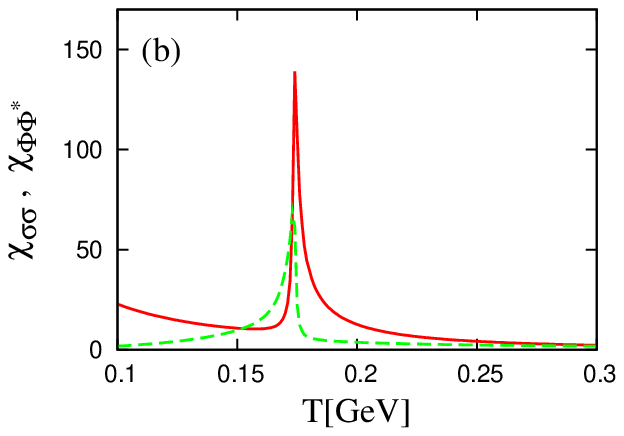}
 \end{center}
\caption{(color online). Results of the EPNJL model for 
$T$ dependence of (a) $\sigma/\sigma_0$ and $\Phi$ 
at $\theta=\Theta =0$ and 
(b) $\chi_{\sigma\sigma}$ and $\chi_{\Phi\Phi^*}$ at $\theta=\Theta =0$. 
The solid (dashed) line represents $\sigma$ ($\Phi$) in panel (a)
 and $\chi_{\sigma\sigma}$ ($\chi_{\Phi\Phi^*}$) in panel (b). 
 In panel (b), $\chi_{\Phi\Phi^*}$ is multiplied by $10^2$. 
}
\label{fig-EPNJL-zero}
\end{figure}

Figure~\ref{fig-EPNJL-zero} shows results of the EPNJL model. 
Panel (a) presents $T$ dependence of $\sigma$ (solid line) and $\Phi$ (dashed line) at $\theta=\Theta =0$, and panel (b) shows 
$T$ dependence of $\chi_{\sigma\sigma}$ (solid line) and $\chi_{\Phi\Phi'}$ (dashed line) at $\theta=\Theta =0$.  
The entanglement vertex \eqref{entanglement-vertex}
makes the chiral and the deconfinement crossover almost coincide.

\begin{figure}[htbp]
\begin{center}
 \includegraphics[width=0.35\textwidth]{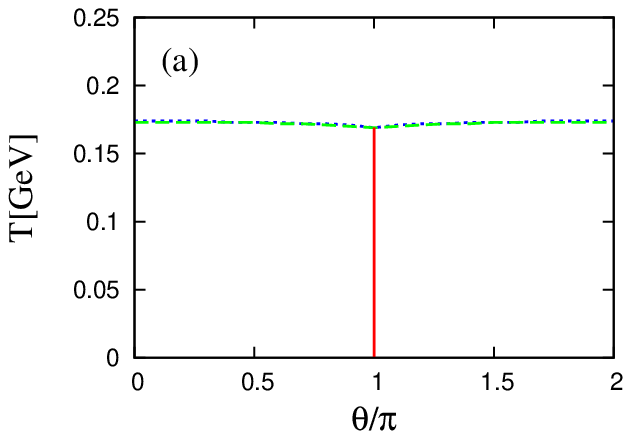}
 \includegraphics[width=0.35\textwidth]{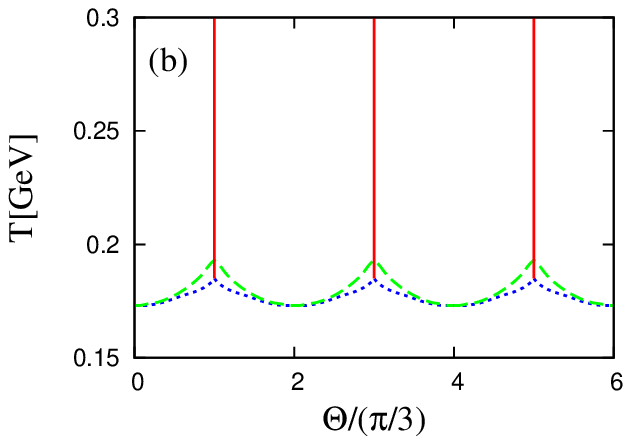}
 \end{center}
\caption{(color online). Phase diagrams in (a) $T$-$\theta$ plane at $\Theta=0$ 
and (b) $T$-$\Theta$ plane at $\theta=0$ in the EPNJL model. 
In panel (a)((b)), the vertical solid line represents the first-order phase transition caused by P violation (C violation). 
The dashed and dotted lines stand for the chiral and the deconfinement crossover, respectively, although they almost coincide. 
}
\label{phase-diagram-EPNJL}
\end{figure}

Figure~\ref{phase-diagram-EPNJL} shows phase diagrams of the EPNJL model. 
Panel (a) presents the phase diagram in $\theta$-$T$ plane at $\Theta =0$, 
and panel (b) shows the phase diagram in $\Theta$-$T$ plane at $\theta =0$. 
In both diagrams we see that curves of the chiral and deconfinement transitions almost coincide. 

\begin{figure}[htbp]
\begin{center}
 \includegraphics[width=0.35\textwidth]{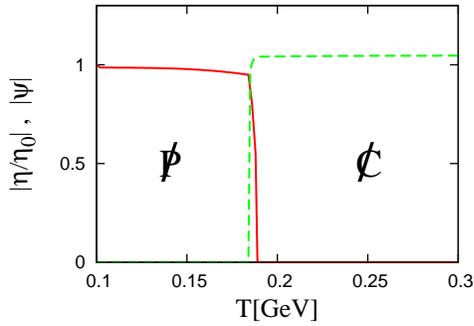}
\end{center}
\caption{(color online) (a) $T$ dependence of  
$|\eta/\eta_0|$ and $|\psi|$ at $\theta =\pi$ and $\Theta =\pi /3$ in the EPNJL model.  
The solid (dashed) line represents $\eta$ ($\psi$). 
}
\label{fig-CP-EPNJL}
\end{figure}

Figure~\ref{fig-CP-EPNJL} shows $T$ dependence of $|\eta/\eta_0|$ and $|\psi|$ 
$\theta=\pi$ and $\Theta =\pi/3$. 
The entanglement vertex also makes P-restoration and C violation 
almost coincide. 
However, since $T_{\rm P}$ is slightly larger than $T_{\rm C}$, there is a narrow region of $T$ where both of P and C symmetries are violated.  

\begin{figure}[htbp]
\begin{center}
 \includegraphics[width=0.35\textwidth]{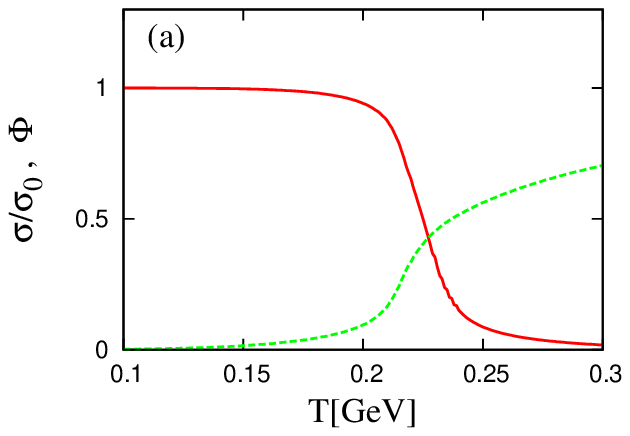}
 \includegraphics[width=0.35\textwidth]{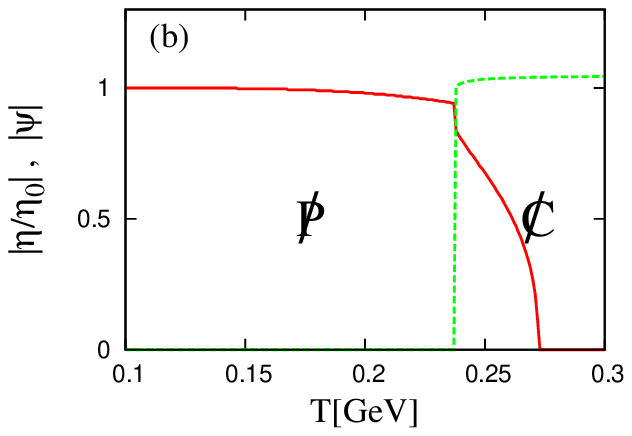}
 \end{center}
\caption{(color online). Results of the PNJL model with $T_0=270$~MeV for 
$T$ dependence of (a) $\sigma/\sigma_0$ and $\Phi$ 
at $\theta=\Theta =0$ 
and (b) $|\eta/\eta_0|$ and $|\psi|$ at $\theta =\pi$ and $\Theta =\pi /3$. 
The solid (dashed) line represents $\sigma$ ($\Phi$) in panel (a) 
and $\eta$ ($\psi$) in panel (b). 
}
\label{fig_PNJL_T0270}
\end{figure}

Here, we set $T_0$ to 270~MeV, the original value extracted from 
the pure-gauge LQCD data. 
Figure \ref{fig_PNJL_T0270}(a) shows $T$ dependence of $\sigma$ and $\Phi$ 
calculated with the PNJL model at $\theta =0$ and $\Theta =0$. 
The difference of the chiral and deconfinement pseudocritical temperatures 
is $T_\chi -T_{\rm d}=232-215$~MeV$=17$~MeV. 
This difference is small compared with the corresponding result in  
the case of $T_0=212$~MeV. 
As seen in Fig. \ref{fig_PNJL_T0270}(b), however, 
for $\theta =\pi$ and $\Theta =\pi/3$ the transition temperature of 
P restoration, $T_{\rm P}=  272$~MeV, 
is larger than that of C violation, $T_{\rm C}=238$~MeV. 
The difference is smaller than the corresponding difference in the case of $T_0=212$~MeV.
For the PNJL model, thus, the transition temperatures, $T_\chi$, $T_{\rm d}$, $T_{\rm P}$ and $T_{\rm C}$, are shifted up by changing $T_0$ from 212~MeV to 270~MeV with about 50~\% reduction of the relative differences, $T_\chi -T_{\rm d}$ and $T_{\rm P}-T_{\rm C}$. 

\begin{figure}[htbp]
\begin{center}
 \includegraphics[width=0.35\textwidth]{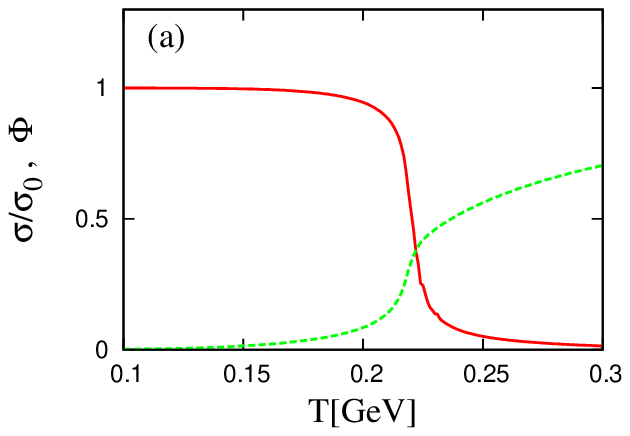}
 \includegraphics[width=0.35\textwidth]{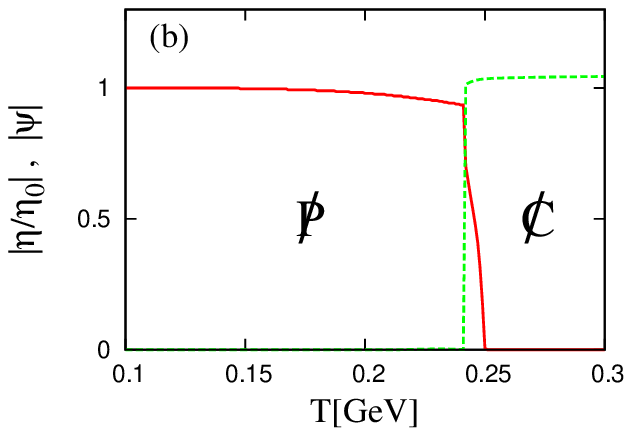}
 \end{center}
\caption{(color online). Results of the EPNJL model with $T_0=270$MeV and $\alpha_1=\alpha_2=0.2$ for 
$T$ dependence of (a) $\sigma/\sigma_0$ and $\Phi$ 
at $\theta=\Theta =0$ 
and (b) $|\eta/\eta_0|$ and $|\psi|$ at $\theta =\pi$ and $\Theta =\pi /3$. 
The solid (dashed) line represents $\sigma$ ($\Phi$) in panel (a) 
and $\eta$ ($\psi$) in panel (b). 
}
\label{fig_EPNJL_T0270}
\end{figure}

Figure \ref{fig_EPNJL_T0270}(a) shows 
$T$ dependence of $\sigma$ and $\Phi$ 
calculated by the EPNJL model with $\a_1=\a_2=0.2$ and $T_0=270$~MeV for 
the case of $\theta =0$ and $\Theta =0$. 
The chiral and deconfinement transitions almost coincide with each other; 
namely, $T_\chi =222$~MeV and $T_{\rm d}=218$~MeV. 
As seen in Fig. \ref{fig_EPNJL_T0270}(b), 
also for $\theta =\pi$ and $\Theta =\pi/3$,  
the transition temperature of P restoration, $T_{\rm P}=249$~MeV, 
almost coincides that of C violation, $T_{\rm C}=242$~MeV. 
For the EPNJL model, thus, 
the transition temperatures, $T_\chi$, $T_{\rm d}$, $T_{\rm P}$ and 
$T_{\rm C}$, are shifted up by changing $T_0$ from 212~MeV to 
270~MeV with keeping the relative differences, 
$T_\chi -T_{\rm d}$ and $T_{\rm P}-T_{\rm C}$, small. 

\begin{figure}[htbp]
\begin{center}
 \includegraphics[width=0.35\textwidth]{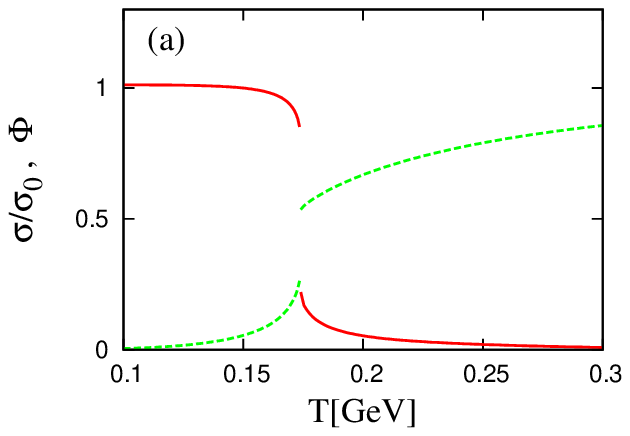}
 \includegraphics[width=0.35\textwidth]{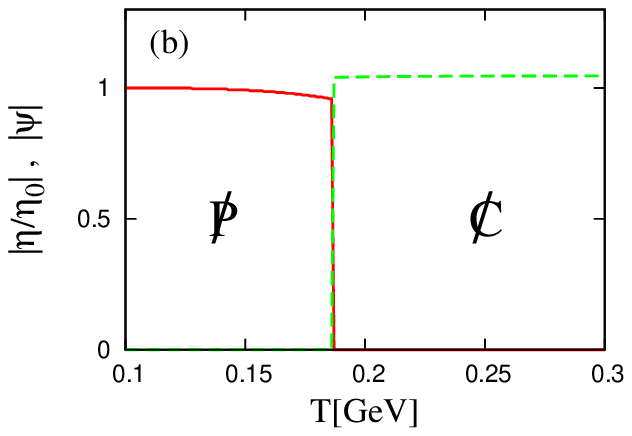}
 \end{center}
\caption{(color online). Results of the EPNJL model with $\alpha_1 =\alpha_2=0.28$ for 
$T$ dependence of (a) $\sigma/\sigma_0$ and $\Phi$ 
at $\theta=\Theta =0$ 
and (b) $|\eta/\eta_0|$ and $|\psi|$ at $\theta =\pi$ and $\Theta =\pi /3$. 
The solid (dashed) line represents $\sigma$ ($\Phi$) in panel (a) 
and $\eta$ ($\psi$) in panel (b). 
}
\label{fig-EPNJL-strong}
\end{figure}

It is widely believed that, in the realistic world, the chiral and the deconfinement transitions are cross over at $\theta =\Theta =0$~\cite{YAoki}. 
However, in a theoretical point of view, it may be interesting to consider both chiral and deconfinement transitions are of first order. 
In fact, in the Holographic QCD, both of the two transitions are of first order~\cite{SS,Aharony}. 
(Recently, the RW transition is also confirmed in the Holographic QCD~\cite{Aarts}. ) 
Figure~\ref{fig-EPNJL-strong}(a) show the $T$ dependence of chiral condensate $\sigma$ and Polyakov-loop $\Phi$ at $\theta =\Theta =0$ when we put $\alpha_1=\alpha_2=0.28$ in the EPNJL model. 
In this case, the chiral restoration and the deconfinement transition are of first order and happen at the same time. 
This result is consistent with Fig. 3 of Ref.~\cite{Gatto:2010pt}. 
Figure~\ref{fig-EPNJL-strong}(b) show the $T$ dependence of $\eta$ condensate and the phase $\psi$ of the modified Polyakov-loop $\Psi$ at $\theta =\theta$ and $\Theta =\pi/3$ with the same parameter set. 
Both the P restoration and the C violation are of first order and perfectly coincide. 
There is only one critical temperature $T_{\rm PC}=$187MeV below (above) which the P(C) is violated. 
In this case, the P and C violations exclude each other perfectly. 
 
\section{Summary}
\label{Summary}

Using the PNJL model, 
we have investigated P-restoration and C violation 
at finite $\theta$ and/or finite $\Theta$, 
where $\theta$ is the parameter of the so-called $\theta$-vacuum, 
$\mu=i\Theta T$ is the imaginary quark-number chemical potential and $T$ is the temperature. 

The P symmetry is spontaneously broken below the critical temperature $T_{\rm P}$ at $\theta =\pi$ and $\Theta =0$, while the C symmetry is spontaneously broken above the critical temperature $T_{\rm C}$ at $\theta=0$ and $\Theta =\pi$. 
The second-order endpoint of P-restoration $\theta =\pi$ 
is connected with the chiral crossover, 
while the first-order endpoint of C violation at $\Theta =\pi$ is done 
with the crossover deconfinement transition. 
As a consequence of these connections, 
the chiral and the deconfinement crossover in the real world with no  
$\theta$ and $\Theta$ turn out be remnants of the endpoint of P-restoration and the triple point of C violation, respectively. 

When $\theta =\pi$ and $\Theta =\pi$ ($\pi/3$, $5\pi/3$), at zero $T$, 
P symmetry is spontaneously broken while C symmetry is conserved. 
As $T$ increases, P symmetry is restored just after C symmetry is 
spontaneously broken, so that either P or C symmetry or 
both the symmetries are spontaneously broken for any $T$. 

Two-flavor LQCD data show that the chiral and the deconfinement crossover 
almost or exactly coincide at $\theta =\Theta=0$. 
The coincidence suggests that P-restoration and C violation at $\theta =\Theta=\pi$ also almost or exactly coincide. 
This suggestion is supported by the EPNJL model that reproduces LQCD data at $\theta =\Theta=0$, at finite imaginary chemical potential and at finite isospin chemical potential. 
It is natural to think that such singular behaviors 
at $\theta =\Theta=\pi$ take place simultaneously. If so, 
the coincidence between P restoration and C violation 
may be an origin of the coincidence between the chiral and the deconfinement 
crossover in the real world with no $\theta$ and $\Theta$. 
Further study along this line is quite interesting. 

~

\begin{acknowledgments}
The authors thank T. Inagaki, A. Nakamura and M. Ruggieri for useful discussions and suggestions. 
H. K. also thanks M. Imachi, H. Yoneyama, M. Matsuzaki, Y. Sasai, N. Isizuka, H. Aoki, T. Saito and M. Tachibana for useful discussions and suggestions. 
This calculation was partially carried out on SX-8 at Research Center for Nuclear physics, Osaka University. 
Y.S. and K.K. are supported respectively by JSPS Research Fellows. 
\end{acknowledgments}

\newpage


\end{document}